\documentclass[aps,prb,superscriptaddress,twocolumn]{revtex4}
\usepackage{amsmath,amssymb,bm}
\usepackage{graphicx}
\usepackage{epstopdf}
\usepackage{latexsym}
\usepackage{subfigure}
\usepackage{color}
\usepackage{amssymb}
\usepackage{wasysym}
\usepackage{verbatim}
\usepackage{makecell,pict2e}

\begin{document}
\date{\today}
\title{Dimer Mott Insulator in an Oxide Heterostructure}

\author{Ru Chen}
\affiliation{Department of Physics, University of California, Santa Barbara, Santa Barbara, CA, 93106}

\author{SungBin Lee}
\affiliation{Department of Physics and Centre for Quantum Materials, University of Toronto, Toronto, Ontario M5S 1A7, Canada}

\author{Leon Balents}
\affiliation{Kavli Institute of Theoretical Physics, University of California, Santa Barbara, Santa Barbara, CA, 93106}

\begin{abstract}
We study the problem of designing an artificial Mott insulator in a correlated oxide heterostructure.  We consider the extreme limit of quantum confinement based on ionic discontinuity doping, and argue that a unique {\em dimer Mott insulator} can be achieved for the case of a single SrO layer in a GdTiO$_3$ matrix.  In the dimer Mott insulator, electrons are localized not to individual atoms but to bonding orbitals on molecular dimers formed across a bilayer of two TiO$_2$ planes, and is analogous to the Mott insulating state of Hubbard ladders, studied in the 1990s.  We verify the existence of the dimer Mott insulator through both {\em ab initio} and model Hamiltonian studies, and find for reasonable values of Hubbard $U$ that it is stable and ferromagnetic with a clear bonding/anti-bonding splitting of order 0.65eV, and a significant smaller Mott gap whose size depends upon $U$.  The combined effects of polar discontinuity, strong structural relaxation and electron correlations all contribute to the realization of this unique ground state.
\end{abstract}

\maketitle

%%%%%%%%%%%%%%%%%
%%%%%%%%%%%%%%%%%

% For decades (half a century?), crystalline transition metal and rare earth compounds, largely oxides, have been the preferred venue to study remarkable phenomena such as Mott metal-insulator transitions, antiferromagnetism, quantum criticality, and superconductivity, arising from the Coulomb interaction between electrons.   In a parallel effort, semiconductor growth and processing has allowed electron-electron interactions to be pursued to the context of two-dimensional heterostructures, which allow exquisite control over the geometry and phenomenal sample purity, but which reveal electron interaction effects only when pushed to extremes of low temperature and high magnetic fields.
Recently, the growth techniques from semiconductor physics, such as Molecular Beam Epitaxy (MBE), have been increasing applied to transition metal and rare earth materials to create “correlated heterostructures”.\cite{mannhart2010oxide}  The resulting atomic layer control promises the ability to design orbital, spin, and charge states, create new emergent phenomena, and study fundamental physics of correlated quantum states in unprecedented new ways.   A first step in this direction would be the creation of the simplest and most dramatic manifestion of electron-electron interactions: the formation of a Mott insulators, a system which would be a metal according to band theory, but in which instead electrons localize because their motion is jammed by their mutual short-range Coulomb repulsion.  % (Nearly all other correlated electron phenomena can be found in close proximity to some Mott insulator.)

Mott insulators occur only when the electron density is commensurate with the underlying lattice, and typically an (odd) integer number of electrons per atom is required.  A charge density of one electron per atom is enormous, reaching of order $n_{2d} \approx 7\times 10^{14} \text{cm}^{-2}$ for a typical perovskite structure even if these electrons are confined to a single atomic layer.  This provides a challenge for heterostructures, as this $n_{2d}$ is already an order of magnitude larger than can be achieved in the highest density semiconductors, and even if it is created, the electron density per atom will be greatly reduced by the electrons’ tendency to spread out in the third dimension.  In this paper, we show that these difficulties can be overcome by judicious design of a Dimer Mott Insulator (DMI), a state envisioned decades ago in the context of one-dimensional Hubbard ladders,\cite{Dagotto02021996} and created here in two dimensions at a single monolayer of SrO embedded in a GdTiO$_3$ matrix.  In the DMI, the requisite high charge density is achieved by combining ionic discontinuity doping, quantum confinement, and the formation of electronic dimers.  The dimers are bonding orbitals on electron pairs, to which electrons are Mott localized instead of to individual atoms.  Dimer formation halves the charge density needed to reach the Mott state, relative to the usual single atom localization, and is crucial to the success of our scheme.  We combine \textit{ab initio} and model calculations to establish the existence and nature of the DMI state theoretically.

The starting point for our work is the polar/ionic discontinuity, which induces a large net charge, typically half an electron per planar unit cell, at a polar to non-polar interface.  % Provided this charge is not compensated by defects or atomic reconstruction, it may appear as itinerant carriers.  
The polar discontinuity has been identified as a possible mechanism of doping in many oxide interface studies,\cite{nakagawa2006some,mannhart2010oxide} but has only recently been quantitatively verified systematically\cite{pouya1}.   In MBE grown heterostructures of GdTiO$_3$ and SrTiO$_3$, a carrier density of $n_{2d}=3.5\times 10^{14} \text{cm}^{-2}$ (=1/2 e$^-$ per planar unit cell) for each GdTiO$_3$/SrTiO$_3$ interface has been systematically observed by Hall coefficient measurements\cite{pouya1}.  These electrons fall into the empty $d$ states of the SrTiO$_3$, and consequently high 3d carrier density can be achieved by confinement in narrow quantum wells of SrTiO$_3$ embedded in thicker GdTiO$_3$, with n$_{3d}$=n$_{2d}$/w, where w is the well width.  Recent transport experiments\cite{pouya2} showed that such wells with a width of a few SrTiO$_3$ unit cells are indeed strongly correlated metals with ultra-high carrier density $n_{2d}=7\times 10^{14} \text{cm}^{-2}$ arising from two interfaces, corresponding to $1/2+1/2=1 e^-$ per planar unit cell.

To maximize the 3d electron density and approach the Mott limit, we take this approach to its logical end and consider the case of a single SrO layer embedded in GdTiO$_3$.  However, even in this case of ultimate confinement, we do not achieve a 3d density of 1 $e^-$ per atom.  This is because the doped electrons go symmetrically into the two “interfaces” – TiO$_2$ layers – on either side of the SrO plane.  While this situation appears unfavorable for the formation of a Mott insulator, all is not lost.  On fundamental grounds, the condition for the formation of a Mott state depends only on the charge per unit cell defined by the translational symmetry of the system.  Here the two interface planes form a bilayer, with translational symmetry only within the plane, and there is indeed a unit charge per planar unit cell.  This indicates that a Mott state is possible in principle.  To realize it, we must somehow induce the two Ti atoms in a unit cell to act as a single `superatom'. % On the other hand, the \textit{ab initio} study shows for two SrO layer embedded case, the interfacial ground state is metallic and there is no existence of DMI (\textbf{put result to appendix.}).

It is instructive to view the bilayer on its side, with the Ti atoms projected into an x-z plane (here we use standard cubic coordinates for the perovskite structure, and z is the growth direction).  The Ti-O-Ti network then forms a “ladder” with Ti-O-Ti bonds between the two interfaces making the rungs of the ladder, and intra-plane bonds projecting to the ladder’s legs.   In a Hubbard model description, symmetry dictates that the hopping amplitude for electrons along the legs, $t$, and along the rungs, $t_\perp$, are unequal.  Correlated electron ladders were studied intensively in the 1990s,\cite{Dagotto02021996} and in particular it was shown when $t_\perp/t$ is sufficiently large (approximately $t_\perp/t > 1$ for the one-dimensional Hubbard ladder), electrons form an unconventional Mott state of bonding orbitals on the rungs of the ladder - the one dimensional analog of the DMI.  Qualitatively, we anticipate the same physics applies to the Hubbard bilayers, provided $t_\perp/t$ is sufficiently large.

%%%
\begin{figure}[t]
  \begin{center}
 \scalebox{0.9}{\includegraphics[width=\columnwidth]{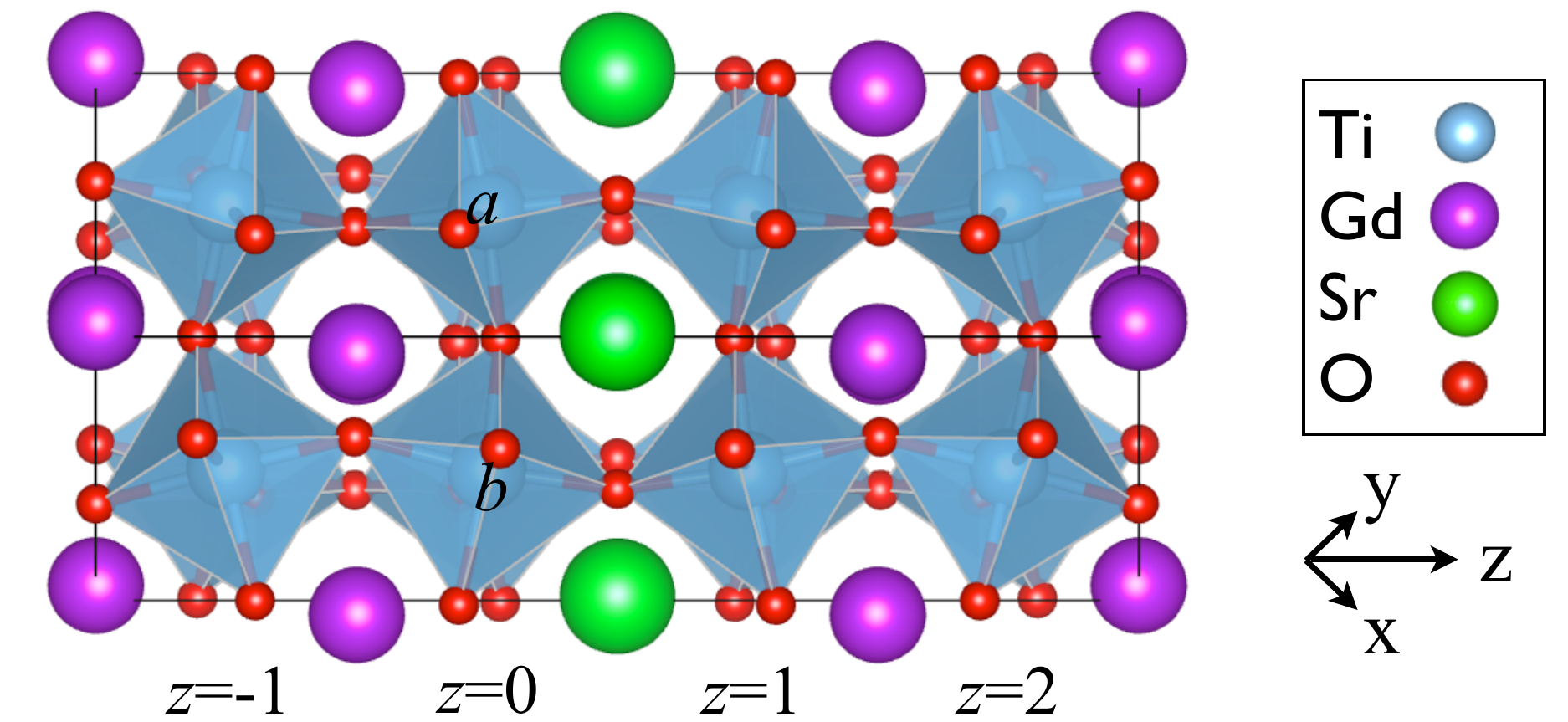}}
  \end{center}
  \caption{(Color online) One unit cell of the relaxed structure of the superlattice (SrTiO$_3$)$_1$(GdTiO$_3$)$_3$. TiO$_6$ octahedra are drawn in blue, to emphasize tilts.. The superlattice repeats periodically along the z direction. The bilayer consists of TiO$_2$ planes at z=0 and z=1. Here $a$ and $b$ denote the two different Ti sublattices.}
      \label{fig:str}
\end{figure}
%%%

Hopping parameters in transition metal oxides are largely controlled by the metal-oxygen-metal bond angle, due to the directionality of $d$ and $p$ orbitals, and are generally largest when the bond angle is closest to 180$^\circ$.  Intuitively, we expect that the interlayer Ti-O-Ti bonds are the most SrTiO$_3$-like, while those within the TiO$_2$ planes conform more closely to those of the GdTiO$_3$.  Since SrTiO$_3$ is nearly perfectly cubic, while GdTiO$_3$ is one of the most highly distorted titanates, this appears quite favorable.  We checked this intuition with \textit{ab initio} density functional theory (DFT) calculations for periodic superlattices of the single SrO layer, (SrTiO$_3$)$_1$(GdTiO$_3$)$_n$, with n=3,5.  Calculations were performed in the Wien2k\cite{wien2k} implementation and the generalized gradient approximation\cite{gga} (GGA). An RKmax parameter 7.0 was chosen with RMTs of 1.91 a.u., 1.69 a.u., 2.29 a.u. and 2.27 a.u. for Ti, O, Sr and Gd, respectively.  The DFT calculation is carried out in a $\sqrt 2a \times \sqrt2 a \times c$ unit cell to allow for the possibility of octahedral tilts, where $a$ is set to be the value of the experimental SrTiO$_3$ lattice constant, 3.905\AA. The structural optimization is done both on the atomic coordinates and $c/a$ ratio, within the GGA + U approximation.   We focus on U$_\text{eff}$=U-J=4eV on the Ti $d$ orbitals, an acceptable value for the titanates\cite{HFfujimori}. In addition, we further add U$_\text{eff}$=U-J=8.5eV on the Gd $f$ orbitals since the energy of the occupied Gd $f$ bands lie much lower than the Fermi energy in practice. This value will not affect the relevant electronic properties. 

%%%
\begin{figure}[t]
  \begin{center}
 \scalebox{0.9}{\includegraphics[width=\columnwidth]{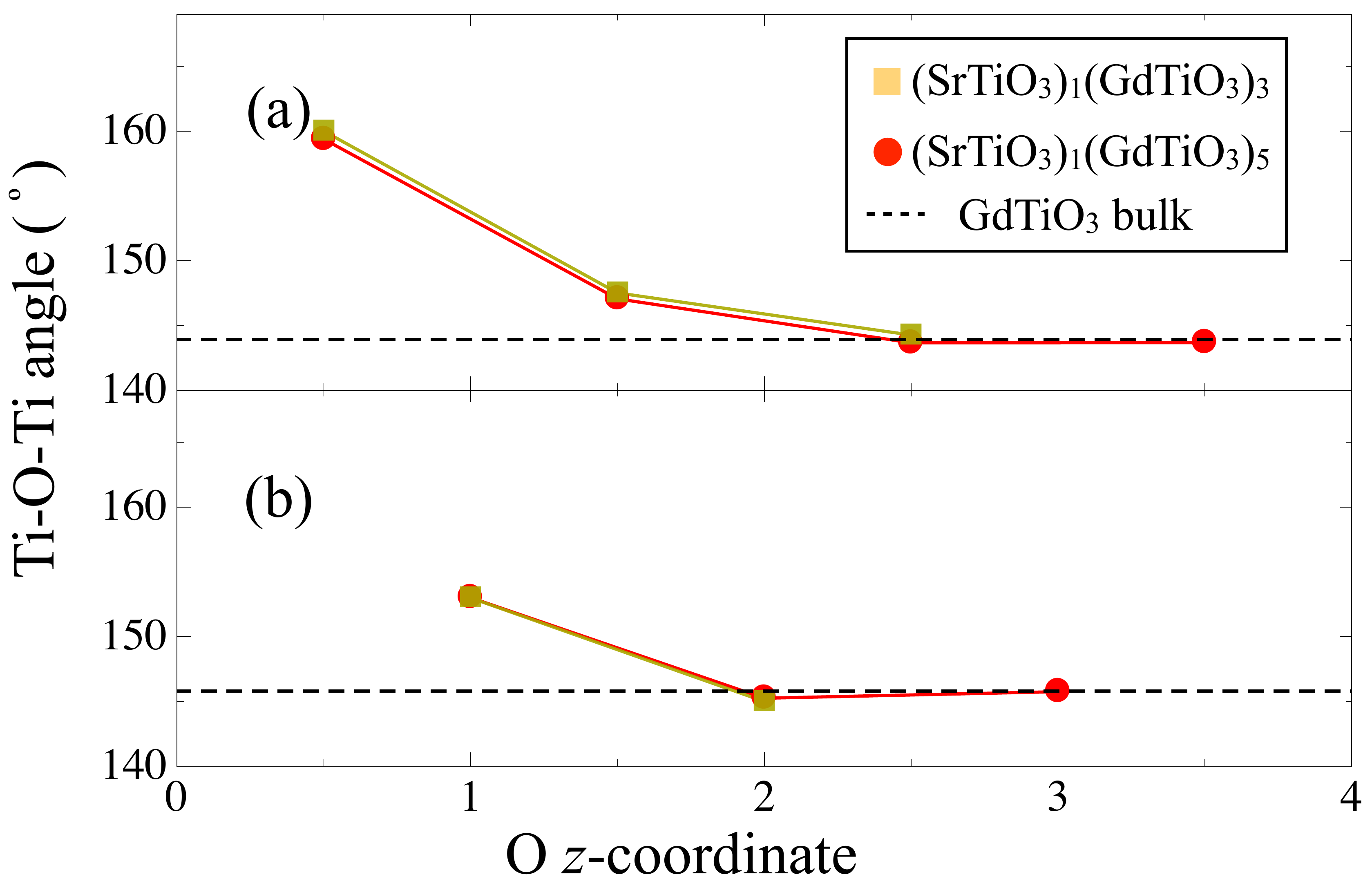}}
  \end{center}
  \caption{(Color online) Ti-O-Ti bond angles along (a) vertical direction and (b) in-plane direction with respect to the O (in the Ti-O-Ti bond) z positions. O along vertical direction resides in the RO (R=Sr, Gd) plane, which is why the coordinate appears to be half-integer. }
      \label{fig:angle}
\end{figure}
%%%

We now discuss the key features of the relaxed structure, which becomes independent of $n$ for $n \geq 3$, and is shown in Fig.~\ref{fig:str}.  First, relaxation is significant at the two interface layers, but decays quickly into the GdTiO$_3$ region. Second, the Ti-O-Ti bond angles are highly direction dependent near the interface, as shown in Fig.~\ref{fig:angle}.  The ``vertical'' Ti-O-Ti bond connecting the two interfaces is slightly distorted, with a 160$^\circ$ angle.   The next vertical Ti-O-Ti bond away from the interface is already highly distorted, with only a 3$^\circ$ angular difference from that of bulk GdTiO$_3$, and more distant bonds are nearly indistinguishable from bulk.   On the other hand, the in-plane bonds are distorted but different from the bulk even in the interfacial layers, with bond angles of about 153$^\circ$.  Third, the Ti-O bond length varies by only 6\% between the longest and shortest bonds in the entire superlattice, and is less significant compared to the dramatic bond angle variations.

%
%%%
\begin{figure}[t]
  \begin{center}
 \scalebox{1.0}{\includegraphics[width=\columnwidth]{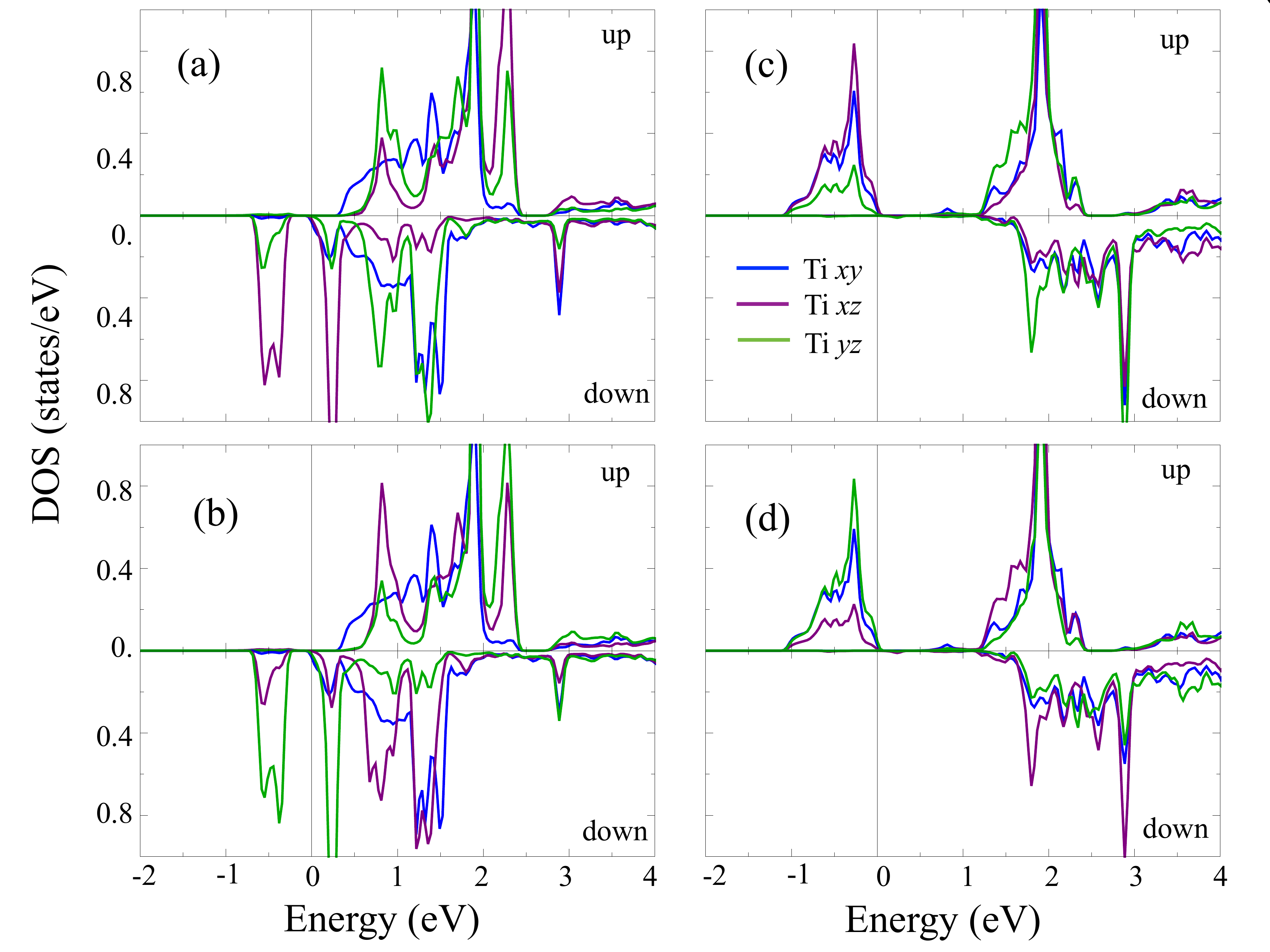}}
  \end{center}
  \caption{(Color online) Layer-resolved electronic density of states of Ti 3$d$ states of (a) sublattice $a$ and of (b) sublattice $b$ at $z=1$ (at the interface); Ti of (c) sublattice $a$ and of (d) sublattice $b$ at $z=2$ (center of GdTiO$_3$ region). The Fermi energy is set to zero.}
      \label{fig:dos}
\end{figure}
%%%
%

Is the bond angle difference, together with other minor structural relaxation effects sufficient to promote a DMI?  We first check the electronic structure within the GGA+U approximation using the relaxed structure.  Searching for possible magnetic structures, we obtained the lowest energy for a state with ferromagnetic alignment of Ti spins within each TiO$_2$ plane, with interfacial Ti spins antiparallel to those in the GdTiO$_3$ region (energy difference compared to parallel to those in the GdTiO$_3$ region is small).  With this configuration, and $U_{\rm eff}=4eV$, the interfacial and bulk density of states (DOS) is shown in Fig.~\ref{fig:dos}.  We observe a bulk gap of 1.25eV, comparable to theoretical values in the literature, but remarkably a much reduced but still non-zero gap at the interface of approximately 0.2eV.  This is a signature of the DMI state.  Within GGA+U, the DMI persists for $U_{\rm eff} \gtrsim 3.5eV$, with both bulk and interfacial gaps reduced for smaller $U$.  To see the DMI more directly, we decomposed the DOS by Ti site and the 3 $t_{2g}$ orbitals (defined by pseudo-cubic axes).   Within the interface plane, the major DOS just below the Fermi level has no $xy$ character, consisting instead of spin down predominantly $xz$/$yz$ orbitals on alternating a/b sublattices.   Consequently, we identify this state as the occupied bonding orbital of the DMI, with the $xz$/$yz$ orbital degeneracy split by octahedral rotations. The anti-bonding state, centered at around 0.2eV, is again mainly composed of $yz$ or $xz$ orbitals.  The separation (of subband centers) from the bonding state gives the bonding/anti-bonding splitting of 0.65eV.   The gap is much smaller than this splitting, however, due the width of the bonding and anti-bonding bands. Away from the interface region, the electronic structure resembles that of bulk GdTiO$_3$.  For the Ti in the center of the GdTiO$_3$ region, we observe dominant $xy$/$yz$ and $xy$/$xz$ states alternating between two orthorhombic sublattices.   Quantitative analysis shows the occupation at $xy$, $yz$ and $xz$ states are 37\%, 16\% and 47\% for Ti at sublattice $a$, similar to in bulk GdTiO$_3$.

To study the DMI state more explicitly, we constructed a model extended Hubbard Hamiltonian by extracting hopping parameters from the \textit{ab initio} calculations. 
Using the optimized structure obtained within GGA+U, a pure GGA calculation was carried out. The eigenvalues of the $f$ states of Gd were shifted manually away from the Fermi energy. We constructed 30 maximally localized Wannier functions\cite{wannier90,wien2wannier} (MLWF) around the Fermi energy, with the 3 $t_{2g}$ orbitals forming the basis. The hopping parameters were then calculated by evaluating the matrix element of the MLWF. Including these hopping parameters, the tight-binding model of the bilayer takes the form
\begin{equation}
\label{eq:htb}
H_{tb}=\sum_{\langle ij \rangle} \sum_{mn,\alpha} t_{ij}^{mn} c_{im\alpha}^\dagger c_{jn\alpha}^{\vphantom\dagger}
\end{equation}
where $i$ and $j$ are nearest neighbor sites, $m$ and $n$ are orbital indices and $\alpha$ is the spin index. The hopping parameters for the two interfacial Ti layers are tabulated in Table~1, which contains the full orbital dependence of the hopping terms.  The maximum intra-plane and inter-plane hopping matrix elements are $t\sim0.35$eV, t$_\perp \sim 0.63$eV, respectively, which gives strong support for the DMI picture. The separation energy between bonding and anti-bonding state is consistent with this magnitude inter-plane hopping.

\begin{table}
\setlength{\tabcolsep}{5.5pt}
\begin{tabular}{c c c|c c}
\hline  \hline
 &\multicolumn{2}{c|}{Direction} & & Direction \\
($im$; $jn$) & [1, 1, 0] & [-1, 1, 0] & ($im$; $jn$) & [0, 0, 1] \\ 
\hline 
(1$a yz$; 1$byz$) & 0 & -0.35 & (0$ayz$; 1$ayz$) & -0.6 \\ 
%\hline 
(1$ayz$; 1$bxz$) & 0.14 & 0.14 & (0$a yz$; 1$a xz$)  & 0 \\ 
%\hline 
(1$axz$; 1$byz$) & -0.12 & -0.12 & (0$a xz$; 1$a yz$) & 0 \\ 
%\hline 
(1$axz$; 1$bxz$) & -0.35 & 0 & (0$a xz$; 1$a xz$) & -0.63 \\ 
%\hline 
(1$a xy$; 1$b xy$) & -0.34 & -0.34 & (0$a xy$; 1$a xy$)  & 0 \\ 
\hline \hline
\end{tabular} 
\label{table:hop}
\caption{Hopping parameters for two TiO$_2$ interface layers from fits to the (SrTiO$_3$)$_1$(GdTiO$_3$)$_3$ superlattice in unit of eV. The index, for example 1$ayz$, stands for sublattice $a$ Ti at $z=1$ with basis $yz$ state. All the major hoppings (larger than 10$\%$ of the largest hopping magnitude) are kept here. }
\end{table}

%\begin{eqnarray}
%H&=&H_{tb}+H_{int}  \\
%H_{int}&=&\sum_i \Big[ U \sum_m n_{im\uparrow} n_{im\downarrow} +U' \sum_{m \neq n} n_{im\uparrow} n_{in\downarrow} \nonumber \\
%&+& \frac{1}{2}(U'-J)\sum_{m \neq n,\alpha} n_{im\alpha} n_{in\alpha}+ \nonumber \\
%&+&J'\sum_{m \neq n} c_{im\uparrow}^\dagger c_{in\downarrow} c_{im\uparrow}^\dagger c_{in\downarrow}\Big]. 
%\end{eqnarray}

We supplement these hopping parameters with interactions,\cite{MITfujimori,HFfujimori} to form the effective Hamiltonian\cite{MITfujimori,HFfujimori}
\begin{align}
\begin{split}
 H&=H_{tb}+H_{int} \\
 H_{int}&=\sum_i \Big[ U \sum_m n_{im\uparrow} n_{im\downarrow} +U' \sum_{m \neq n} n_{im\uparrow} n_{in\downarrow} \\
  &+ \frac{1}{2}(U'-J)\sum_{m \neq n,\alpha} n_{im\alpha} n_{in\alpha}+J\sum_{m \neq n} c_{im\uparrow}^\dagger c_{in\uparrow} c_{in\downarrow}^\dagger c_{im\downarrow}  \\
&+J'\sum_{m \neq n} c_{im\uparrow}^\dagger c_{in\uparrow} c_{im\downarrow}^\dagger c_{in\downarrow}\Big], 
\end{split}
\end{align}
where $U$ and $U'$ represent on-site intra-orbital and inter-orbital Coulomb repulsion between up and down spin, respectively, and J is the Hund coupling. We will not restrict the condition to Slater-Kanamori interaction parameter $U'=U-2J$ and $J=J'$, but rather simply assume $J=J'$ and explore the phase diagram by varying all the other parameters.

%%%
\begin{figure}[t]
  \begin{center}
    \scalebox{0.9}{\includegraphics[width=\columnwidth]{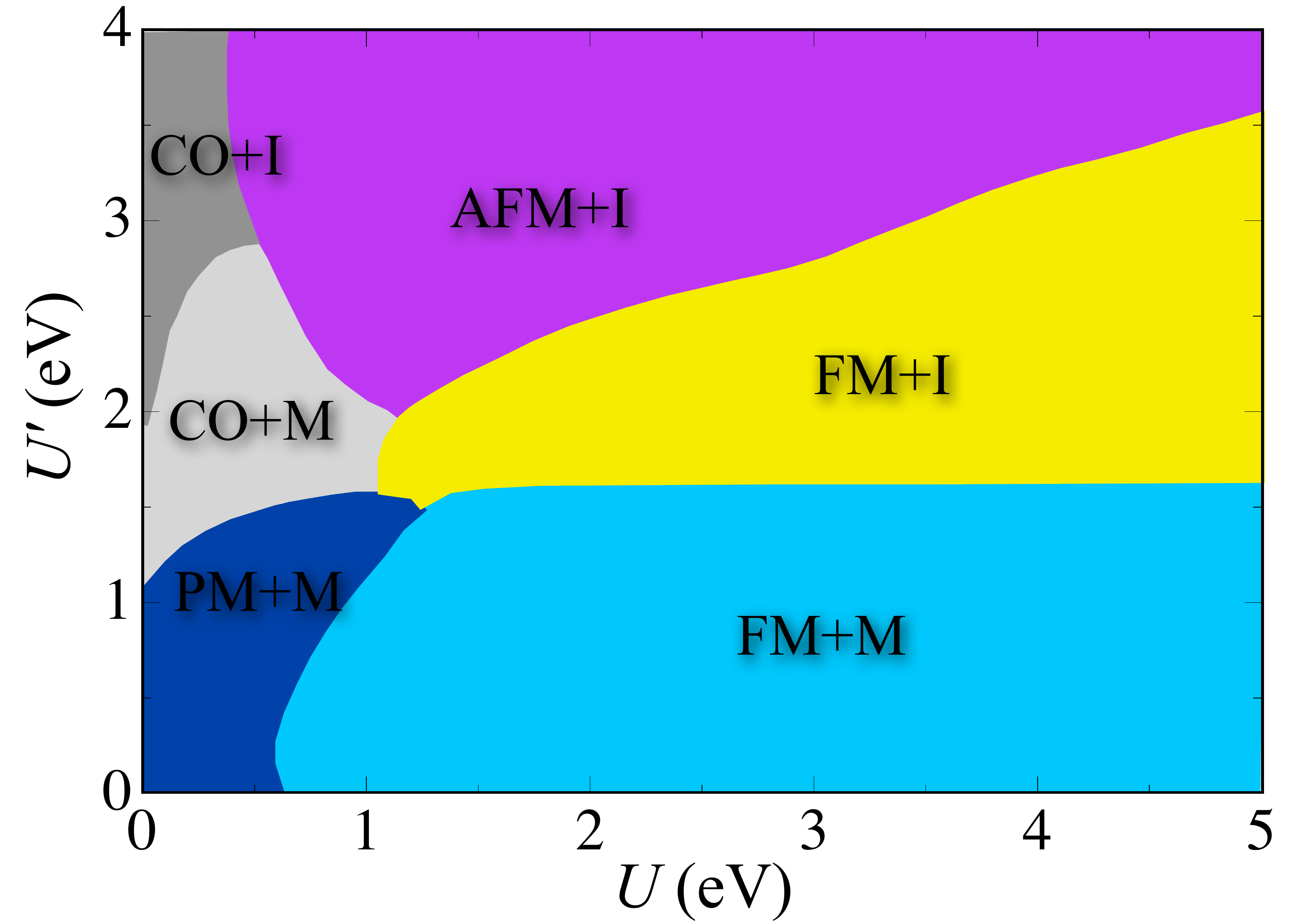}}
  \end{center}
  \caption{(Color online) Hartree-Fock phase diagram for Hund's couping $J=$0.6eV. Here, we abbreviate the phases as follows: PM+M = paramagnetic metal; CO+M= Weakly charge ordered metal; CO+I = charge ordered insulator; FM+M = ferromagnetic metal; FM+I = ferromagnetic insulator; AFM+I = antiferromagnetic insulator.}
      \label{fig:phase}
\end{figure}
%%%

Fig.~\ref{fig:phase} shows the phase diagram as a function of $U$ and $U'$ obtained by the Hartree-Fock approximation, with fixed Hund's coupling $J=$0.6eV. % The Hund's coupling, although usually small, tends to maximize the site spin, thus reduces the inter-orbital Coulomb repulsion between the same spin.
When the Coulomb repulsion is small, the ground state is just paramagnetic metal.  In the unphysical region where inter-orbital repulsion is dominant, $U'>U$, electrons reside in a single orbital per Ti and in fact charge order.  Of most interest is the lower right part of the phase diagram, where $U>U'$. In this region, ferromagnetism arises for sufficiently large $U$. In the weak inter-orbital repulsion limit, the electrons are distributed in all three t$_{2g}$ orbitals.  Under this condition, the system is always metallic or semi-metallic, since those electrons in $xy$ orbitals are non-bonding between layers and hence metallic.  For sufficiently large $U'$, however, the inter-orbital repulsion eventually disfavors and empties the $xy$ states, in favor of the $xz$/$yz$ orbitals which have lower energy through inter-layer hopping.  In this way the bonding state becomes fully occupied and a (ferromagnetic) DMI is achieved.  Further increase of $U'$ enhances the gap and eventually prefers an antiferromagnetic DMI, since Hund's coupling becomes ineffective if there is strictly one electron per site, and super-exchange becomes dominant. For completeness, we also have tested $J$=0,1eV, which shows the Hund's coupling enhances the ferromagnetic insulating state.  For our best guess at physically appropriate values, e.g. $U=4.5$eV and the Slater-Kanamori $U'=U-2J$, the ferromagnetic DMI obtains.   The Hartree-Fock results for these values for the band gap and orbital ordering are very similar to those of the previously discussed GGA+U calculations.   % We note that the intra-orbital repulsion $U’$ plays the important role here of depopulating the $xy$ orbitals, which is important for the DMI formation.

In summary, we have argued for the existence of a Dimer Mott Insulator (DMI) for a single SrO layer embedded in a thick GdTiO$_3$ matrix, using both {\em ab initio} and model calculations.   The DMI state is unique to the bilayer TiO$_2$ structure created by a single SrO layers;  we have indeed verified that a metallic state is obtained in GGA+U for the case of two SrO layers embedded in GdTiO$_3$ (see supplemental material).  Insulating behavior in such structures experimentally should therefore be attributed to the combined effects of disorder (e.g. SrTiO$_3$ thickness fluctuations) and interactions.   The DMI for a single SrO layer could be experimentally probed by many experiments, including transport, optical measurements of the gap and bonding-antibonding splitting, and angle-resolved photoemission.  Observing the magnetic structure is more difficult, but might be possible with ferromagnetic resonance or optical dichroism.  This work suggests many directions for future theoretical and experimental research. We anticipate that the Mott state can be controlled and modified by varying composition, strain and the growth direction.    It may be possible to create antiferromagnetic DMIs by varying the rare earth ion; however, theory is needed to gauge whether this also may destabilize the dimer formation itself.  Choice of substrate also effects the strain and growth direction of the titanate films.   Whether the DMI persists when the GdTiO$_3$ grows along the (110) direction is an important question for future study.  More speculatively, we might contemplate the possibility of superconductivity induced by doping the DMI, by analogy to the superconductivity predicted theoretically and observed experimentally in ladder systems.

We thank Jim Allen, Susanne Stemmer, Dan Ouellette, Pouya Moetakef and Chuck-Hou Yee for helpful discussions. 
We acknowledge support from the Center for Scientific Computing at UCSB: NSF CNS-0960316. This research is supported by DARPA grant No.W911-NF-12-1-0574 (L.B. and R.C.) and the  MRSEC Program of the National Science Foundation, Award No. DMR 1121053, NSERC, CIFAR (SB.L.).

\bibliography{Biblio}

\begin{thebibliography}{11}
\expandafter\ifx\csname natexlab\endcsname\relax\def\natexlab#1{#1}\fi
\expandafter\ifx\csname bibnamefont\endcsname\relax
  \def\bibnamefont#1{#1}\fi
\expandafter\ifx\csname bibfnamefont\endcsname\relax
  \def\bibfnamefont#1{#1}\fi
\expandafter\ifx\csname citenamefont\endcsname\relax
  \def\citenamefont#1{#1}\fi
\expandafter\ifx\csname url\endcsname\relax
  \def\url#1{\texttt{#1}}\fi
\expandafter\ifx\csname urlprefix\endcsname\relax\def\urlprefix{URL }\fi
\providecommand{\bibinfo}[2]{#2}
\providecommand{\eprint}[2][]{\url{#2}}

\bibitem[{\citenamefont{Mannhart and Schlom}(2010)}]{mannhart2010oxide}
\bibinfo{author}{\bibfnamefont{J.}~\bibnamefont{Mannhart}} \bibnamefont{and}
  \bibinfo{author}{\bibfnamefont{D.}~\bibnamefont{Schlom}},
  \bibinfo{journal}{Science} \textbf{\bibinfo{volume}{327}},
  \bibinfo{pages}{1607} (\bibinfo{year}{2010}).

\bibitem[{\citenamefont{Dagotto and Rice}(1996)}]{Dagotto02021996}
\bibinfo{author}{\bibfnamefont{E.}~\bibnamefont{Dagotto}} \bibnamefont{and}
  \bibinfo{author}{\bibfnamefont{T.~M.} \bibnamefont{Rice}},
  \bibinfo{journal}{Science} \textbf{\bibinfo{volume}{271}},
  \bibinfo{pages}{618} (\bibinfo{year}{1996}).

\bibitem[{\citenamefont{Nakagawa et~al.}(2006)\citenamefont{Nakagawa, Hwang,
  and Muller}}]{nakagawa2006some}
\bibinfo{author}{\bibfnamefont{N.}~\bibnamefont{Nakagawa}},
  \bibinfo{author}{\bibfnamefont{H.}~\bibnamefont{Hwang}}, \bibnamefont{and}
  \bibinfo{author}{\bibfnamefont{D.}~\bibnamefont{Muller}},
  \bibinfo{journal}{Nature materials} \textbf{\bibinfo{volume}{5}},
  \bibinfo{pages}{204} (\bibinfo{year}{2006}).

\bibitem[{\citenamefont{Moetakef et~al.}(2011)\citenamefont{Moetakef, Cain,
  Ouellette, Zhang, Klenov, Janotti, Van~de Walle, Rajan, Allen, and
  Stemmer}}]{pouya1}
\bibinfo{author}{\bibfnamefont{P.}~\bibnamefont{Moetakef}},
  \bibinfo{author}{\bibfnamefont{T.~A.} \bibnamefont{Cain}},
  \bibinfo{author}{\bibfnamefont{D.~G.} \bibnamefont{Ouellette}},
  \bibinfo{author}{\bibfnamefont{J.~Y.} \bibnamefont{Zhang}},
  \bibinfo{author}{\bibfnamefont{D.~O.} \bibnamefont{Klenov}},
  \bibinfo{author}{\bibfnamefont{A.}~\bibnamefont{Janotti}},
  \bibinfo{author}{\bibfnamefont{C.~G.} \bibnamefont{Van~de Walle}},
  \bibinfo{author}{\bibfnamefont{S.}~\bibnamefont{Rajan}},
  \bibinfo{author}{\bibfnamefont{S.~J.} \bibnamefont{Allen}}, \bibnamefont{and}
  \bibinfo{author}{\bibfnamefont{S.}~\bibnamefont{Stemmer}},
  \bibinfo{journal}{Appl. Phys. Lett.} \textbf{\bibinfo{volume}{99}},
  \bibinfo{pages}{232116} (\bibinfo{year}{2011}).

\bibitem[{\citenamefont{Moetakef et~al.}(2012)\citenamefont{Moetakef, Jackson,
  Hwang, Balents, Allen, and Stemmer}}]{pouya2}
\bibinfo{author}{\bibfnamefont{P.}~\bibnamefont{Moetakef}},
  \bibinfo{author}{\bibfnamefont{C.~A.} \bibnamefont{Jackson}},
  \bibinfo{author}{\bibfnamefont{J.}~\bibnamefont{Hwang}},
  \bibinfo{author}{\bibfnamefont{L.}~\bibnamefont{Balents}},
  \bibinfo{author}{\bibfnamefont{S.~J.} \bibnamefont{Allen}}, \bibnamefont{and}
  \bibinfo{author}{\bibfnamefont{S.}~\bibnamefont{Stemmer}},
  \bibinfo{journal}{Phys. Rev. B} \textbf{\bibinfo{volume}{86}},
  \bibinfo{pages}{201102} (\bibinfo{year}{2012}).

\bibitem[{\citenamefont{Blaha et~al.}(2001)\citenamefont{Blaha, Schwarz,
  Madsen, Kvasnicka, and Luitz}}]{wien2k}
\bibinfo{author}{\bibfnamefont{P.}~\bibnamefont{Blaha}},
  \bibinfo{author}{\bibfnamefont{K.}~\bibnamefont{Schwarz}},
  \bibinfo{author}{\bibfnamefont{G.~K.~H.} \bibnamefont{Madsen}},
  \bibinfo{author}{\bibfnamefont{D.}~\bibnamefont{Kvasnicka}},
  \bibnamefont{and} \bibinfo{author}{\bibfnamefont{J.}~\bibnamefont{Luitz}},
  \emph{\bibinfo{title}{WIEN2k, An Augmented Plane Wave Plus Local Orbitals
  Program for Calculating Crystal Properties}} (\bibinfo{publisher}{Vienna
  University of Technology, Austria}, \bibinfo{year}{2001}).

\bibitem[{\citenamefont{Perdew et~al.}(1996)\citenamefont{Perdew, Burke, and
  Ernzerhof}}]{gga}
\bibinfo{author}{\bibfnamefont{J.~P.} \bibnamefont{Perdew}},
  \bibinfo{author}{\bibfnamefont{K.}~\bibnamefont{Burke}}, \bibnamefont{and}
  \bibinfo{author}{\bibfnamefont{M.}~\bibnamefont{Ernzerhof}},
  \bibinfo{journal}{Phys. Rev. Lett.} \textbf{\bibinfo{volume}{77}},
  \bibinfo{pages}{3865} (\bibinfo{year}{1996}).

\bibitem[{\citenamefont{Mizokawa and Fujimori}(1996)}]{HFfujimori}
\bibinfo{author}{\bibfnamefont{T.}~\bibnamefont{Mizokawa}} \bibnamefont{and}
  \bibinfo{author}{\bibfnamefont{A.}~\bibnamefont{Fujimori}},
  \bibinfo{journal}{Phys. Rev. B} \textbf{\bibinfo{volume}{54}},
  \bibinfo{pages}{5368} (\bibinfo{year}{1996}).

\bibitem[{\citenamefont{Mostofi et~al.}(2008)\citenamefont{Mostofi, Yates, Lee,
  Souza, Vanderbilt, and Marzari}}]{wannier90}
\bibinfo{author}{\bibfnamefont{A.}~\bibnamefont{Mostofi}},
  \bibinfo{author}{\bibfnamefont{J.}~\bibnamefont{Yates}},
  \bibinfo{author}{\bibfnamefont{Y.}~\bibnamefont{Lee}},
  \bibinfo{author}{\bibfnamefont{I.}~\bibnamefont{Souza}},
  \bibinfo{author}{\bibfnamefont{D.}~\bibnamefont{Vanderbilt}},
  \bibnamefont{and} \bibinfo{author}{\bibfnamefont{N.}~\bibnamefont{Marzari}},
  \bibinfo{journal}{Comp. Phys. Comm.} \textbf{\bibinfo{volume}{178}},
  \bibinfo{pages}{685} (\bibinfo{year}{2008}).

\bibitem[{\citenamefont{Kune{\v{s}} et~al.}(2010)\citenamefont{Kune{\v{s}},
  Arita, Wissgott, Toschi, Ikeda, and Held}}]{wien2wannier}
\bibinfo{author}{\bibfnamefont{J.}~\bibnamefont{Kune{\v{s}}}},
  \bibinfo{author}{\bibfnamefont{R.}~\bibnamefont{Arita}},
  \bibinfo{author}{\bibfnamefont{P.}~\bibnamefont{Wissgott}},
  \bibinfo{author}{\bibfnamefont{A.}~\bibnamefont{Toschi}},
  \bibinfo{author}{\bibfnamefont{H.}~\bibnamefont{Ikeda}}, \bibnamefont{and}
  \bibinfo{author}{\bibfnamefont{K.}~\bibnamefont{Held}},
  \bibinfo{journal}{Comp. Phys. Comm.} \textbf{\bibinfo{volume}{181}},
  \bibinfo{pages}{1888} (\bibinfo{year}{2010}).

\bibitem[{\citenamefont{Imada et~al.}(1998)\citenamefont{Imada, Fujimori, and
  Tokura}}]{MITfujimori}
\bibinfo{author}{\bibfnamefont{M.}~\bibnamefont{Imada}},
  \bibinfo{author}{\bibfnamefont{A.}~\bibnamefont{Fujimori}}, \bibnamefont{and}
  \bibinfo{author}{\bibfnamefont{Y.}~\bibnamefont{Tokura}},
  \bibinfo{journal}{Rev. Mod. Phys.} \textbf{\bibinfo{volume}{70}},
  \bibinfo{pages}{1039} (\bibinfo{year}{1998}).

\end{thebibliography}

%%

%%%
%%%%
%\begin{figure}[t]
%  \begin{center}
% \scalebox{0.95}{\includegraphics[width=\columnwidth]{bond-angle.pdf}}
%  \end{center}
%  \caption{Layer-resolved Ti-O-Ti bonding angles along (a) [001] direction and (b) [110] or [1-10] direction. Both graphs only show the angle for half a unit cell. The x value is defined as what plane the O sits for a Ti-O-Ti angle. l and l-1 are both the interface TiO$_2$ planes. Maybe showing only the 1-4 superlattice rather than all of them.}
%      \label{fig:angle}
%\end{figure}
%%%%
%%%

%\begin{equation}
%\begin{array}{cccc}
%[110] & dyz & dxz & dxy \\ 
%dyz & -0.18 & 0.08 & 0 \\ 
%dxz & -0.08 & -0.18 & 0 \\ 
%dxy & 0 & 0 & -0.17
%\end{array} 
%\end{equation}

\end{document}